# A decentralized trust-aware collaborative filtering recommender system based on weighted items for social tagging systems

Hossein Monshizadeh Naeen[1,2], Mehrdad Jalali[3]

[1] Department of Computer Engineering Neyshabur Branch, Islamic Azad University, Neyshabur, Iran, monshizadeh@iau-neyshabur.ac.ir

[2] Faculty of Computer and Information Technology Engineering, Qazvin Branch, Islamic Azad University, Qazvin,Iran, monshizadeh@qiau.ac.ir

[3] Department of Computer Engineering, Islamic Azad University of Mashhad, Mashhad, Iran, jalali@mshdiau.ac.ir

*Abstract*

*Recommender systems are used with the purpose of suggesting contents and resources for users in a social network. These systems use ranks or tags each user assign to different resources to predict or make suggestions to users. Lately, social tagging systems, in which users can insert new contents, tag, organize, share, and search for contents are becoming more popular. These systems have a lot of valuable information, but data growth is one of its biggest challenges and this has led to the need for recommender systems that will predict what each user may like or need. One approach to the design of these systems which uses social environment of users is known as collaborative filtering (CF). One of the problems in CF systems is trustworthy of users and their tags. In this work, we consider a trust metric (which is concluded from users tagging behavior) beside the similarities to give suggestions and examine its effect on results. On the other hand, a decentralized approach is introduced which calculates similarity and trust relationships between users in a distributed manner. This causes the capability of implementing the proposed approach among all types of users with respect to different types of items, which are accessed by unique id across heterogeneous networks and environments. Finally, we show that the proposed model for calculating similarities between users reduces the size of the user-item matrix and considering trust in collaborative systems can lead to a better performance in generating suggestions.*

## 1. Introduction

Social networks and multimedia content sharing web sites have become increasingly popular in recent years [1], and then recommender systems (RS) have emerged as an important response to the so-called information overload problem in which users are finding it difficult to locate the right information at the right time [2]. These systems are applied in a variety of applications and fields. Social tagging systems such as Delicious, CiteULike, Flickr, etc., allow users to assign personal labels to resources based on their own background knowledge with a purpose to manage, organize, share, discover and recover resources, utilize recommender systems for their suggestion and prediction process [3]. In different types of RSs, they aggregate users' reviews of products, services, and each other, into valuable information. Notable commercial examples include Amazon, Google's page ranking system [e.g. 4], and the Epinions web of trust/reputation system [e.g. 5]. One of RS implementation models which we use in this work are known as Collaborative Filtering (CF) recommender systems. CFs have some weaknesses; data sparsity is the first serious issue of collaborative filtering, i.e. the most active users will only have tagged or rated a small subset of the overall database. Thus, even the most popular items have been viewed by very few users. This weakness is especially evident on cold start users, users who are new in the system or we have very little existing data about them, so we have problems providing good recommendations to these users. The other weakness is Scalability problem. Usually there are millions of users and products. Thus, a large amount of computation power is often necessary to calculate recommendations. The last weakness we mention here is noisy or spam tags, users may make mistakes when they are using the system, or CF results can easily be affected by malicious users' attacks or some agents give wrong tags on purpose for advertisement [6].

In this article in general, we use collaborative filtering, and we try to give more accurate recommendations to users, in particular. For this we assign a degree of trust to each user, according to their tagging behavior in the network. Some neighbors of a user, who are similar to them, based on the resources they have tagged, might not be trustworthy enough to participate in recommendation process. Briefly, Similarity metric on its own is not reliable, so beside similarity of a user, credibility should be considered too. To compute trust values, we study user's behavior in the system based on tags they assigned to different resources and comparing their view with global views.

The remainder of the paper is organized as follows. In Section 2 we introduce the background of recommender systems. In Section 3 we define the problem. We describe proposed model Section 4. Section 5 includes results and evaluation. In Section 6 we talk about conclusions and future works.

## 2. Background

As the number of recommender systems is increasing and more users are getting interested in social networking sites, and also because of information explosion in these systems, many approaches on recommender systems have been studied and proposed. In this section we will review related works on recommender systems with tag orientation and also the studies that are done in the trust field.

*2.1. Context-based studies*

There have already been a reasonable amount of researches in using different attributes that exist in social networks as background knowledge to give recommendation in RS such as [7,8,9], but none of them did not use tags as an attribute to decide on predictions. Some other studies work on using tagging information for recommendation purposes based on the recommendation of tags for assisting the user in annotation related tasks, or to predict tags for contents of the site[10,11]. But Contextual information has been proved to be efficient in recommender systems. Adomavicius et al. [12] presented a reduction-based approach for multidimensional recommendation model that incorporates contextual information, such as place and time, Companion, into the process of recommendation. Vig et al. [13] used tags given to movies of MovieLens site and defined two different metrics, tag relevance and tag preference. tag relevance describes the relationship between a tag and an item, and tag preference shows the user's sentiment towards a tag. They give a list of suggested movies to each client, using basic CF based on ratings users give to movies. Then they show relevance and preference scales to help the user understand why an item was recommended (justification) and also to help users make good decisions by this information.

*2.2. Tag-based CF*

Interests in involving tags in collaborative filtering recommender systems have been increased in recent years. Tso-Sutter et al.[14] proposed an approach to integrate tags in recommender systems by first extending the user-item matrix and then applying an algorithm that fuses two popular RS algorithms, item based CF and user based CF, such that the correlations between users, items and tags can be captured simultaneously. But they have not mentioned the scalability in their work, as long as they extend matrix in two dimensions, it seems the scalability problem gets much worse in this case. Nakamoto et al. [15] introduced a tag based CF which uses tags overlaps to compute similarities between users in two different approaches. However their work may have problem if there are not sufficient overlaps of tags. In another work, Zheng et al. [16] worked on a tag and trust based recommender system, in which they try to get better results by giving weight to temporal information and then combine the time weights with similarity weights.

*2.2. Trust-based CF*

In trust field there are different works done according to information and facilities that exist in the social network which we will review some of them in the follow. O'Donovan et al. [17] proposed a number of computational models of trust based on the past rating behavior of individual profiles, these models operate at the profile-level (average trust for the profile overall) and at the profile-item-level (average trust for a particular profile when it comes to making recommendations for a specific item) and then they discuss about 3 different ways to incorporate these trust values into the recommendation process, and the results show that prediction error has been reduced when trust is considered, by the way their approach is only based on ratings and does not involve tags. In other article [18] users can review items (such as cars, books, movies, software, etc.) and also assign them numeric ratings in the range 1 (min) to 5 (max). Users can also express their Web of Trust, i.e. reviewers whose reviews and ratings they have consistently found to be valuable or not valuable, then they aggregate all the trust statements to produce a trust network then to interfere trust values between users who have not directly declare their trust to other users (trust propagation). Their work as you see, considers trust and also covers cold start problem so can calculate similarity between more users, even new users. But it needs users in the system declare their degree of trust to other users, which is not supported by all social networking sites. Jin et al. [19] have used similarity of friends as a trust metric, and then defined a modified random walk algorithm on the friends trust network to recommend top-k items to each user. Their work needs a social network with the ability to add users as friends. Noll et al. [20] introduced the time of tagging as an additional dimension for assessing the trust of a user in Delicious. They proposed a graph-based algorithm, called spammingresistant expertise analysis and ranking (SPEAR). It computes the expertise score of a user and the quality score of a content which are dependent on each other. The time of tagging is considered so that the earlier a user tags a content, the more expertise score he/she receives. These two scores are calculated iteratively in a similar way to that of the HITS algorithm. It was shown that SPEAR produces better ranking of users than the HITS method. SPEAR was able to demote different types of spammers and remove them from the top of the ranking.

## 3.    Problem definition

*3.1. Collaborative Filtering Systems*

Recommender systems (RS) predict ratings of items or suggest a list of items that is unknown to the user[14]. They take the users, items as well as the ratings or tags of items into account. In general there are two types of Recommender Systems, content-based and collaborative filtering [18]. Content-Based approaches use the characteristics of items and try to recommend items that are similar to those that a user liked in the past[20]. Collaborative Filtering considers users' social environment i.e. they collect and analyzing a large amount of information on users' behaviors, activities or preferences in the social network and predict what each user will like. The underlying assumption of the collaborative filtering approach is that if a person A has the same opinion as a person B on an issue, A is more likely to have B's opinion on a different issue [21]. Collaborative filtering has been widely used in applications which produce a predicted likeliness score or a list of top-N recommended resources for a given user [22]. User-item matrix acts as a basic part in CF, user similarity are discovered from matrix and resource recommendation is also calculated based on the matrix [16]. In rating based systems values of the matrix are fulfilled with the user given ratings, but in tag based approaches traditional log-based ratings are generated based on user's log behaviors with each element taken a binary value 1 or 0, indicating whether a user has viewed a resource [23]. In this case, each viewed resource plays an equal role for a user and ratings are static over time, which may be against the common sense that people tend to present different preferences towards different resources. As we will explain later in this paper we estimate a degree of users interests to each different item that they tag rather than binary values.

*3.2. Trust*

Across most current research, definitions of trust fall into various categories, and a solid definition for it, in many cases can be quite elusive [17], because there are different trust views between users. There are algorithms whose goal is to predict, based on the trust network, the trustworthiness of "unknown" users, i.e. users in which a certain user didn't express a trust statement [18]. There are many different trust metrics, an important classification of trust metrics is global and local ones [24]. Local trust metrics take into account the very personal and subjective views of the users and predict different values of trust in other users for every single user. Instead global trust metrics predict a global "reputation" value that approximates how the community as a whole considers a certain user. From another point of view in a social tagging system, spam or noise can be injected at three different levels: spam content, spam tag-content association, and spammer [26]. Trust modeling can be performed at each level separately or different levels can be considered jointly to produce trust models, for example, to assess a user's reliability, one can consider not only the user profile, but also the content that the user uploaded to a social system (e.g., [27]). So from this aspect trust can get categorized into two classes according to the target of trust, i.e., user and content trust modeling[1]

The distribution of the trust values of the users or contents in a social network can be used to represent the health of that network. Content trust modeling is used to classify content (e.g., Web pages, images, and videos) as spam or legitimate. In this case, the target of trust is a content (resource), and thus a trust score is given to each content based on its content and/or associated tags. In user trust modeling, trust is given to each user based on the information extracted from a user's account, their interaction with other participants within the social network, and/or the relationship between the content and tags that the user contributed to the social tagging system [28]. There are other classifications for trust, for example a trust value can be static, it means trust values are given by users or an administrator assign trust values to the clients, or it can be dynamic, i.e., a user's trust in a social tagging system is dynamic and it changes over time according to the user behavior in the social environment, our proposed model which is explained in the next section is based on a dynamic trust model, in which we conclude users' trust values from the transactions they have had in the system.

## 4. Proposed model

Our goal is to introduce a model for collaborative filtering systems which can find similarities and can create trust relationships among all types of similar users, with respect to different types of items, accessible through unique id (e.g. URL) across heterogeneous social tagging systems. Fig.1. shows the proposed model scheme. For this purpose our system is made up of 4 main modules. The first module is profile manager, which either creates or updates profiles of users and items. The second module will find the similarities between users by an item-weighing strategy in a distributed manner. The third part estimates trust between a user and its neighbors, also done in a decentralized manner. Finally the recommendation module selects k neighbors with highest ranks (i.e. with high similarity and trust values) and makes suggestions. Note that there is no global view of a trust network for users and they are only provided with information regarding their neighbors and their tagging history. Therefore, it is possible to maintain user and item profiles in different groups on several servers to achieve better scalability. To cope with privacy requirements, these servers can be located in different organizations, while profiles of users and items are accessible only through their URL.

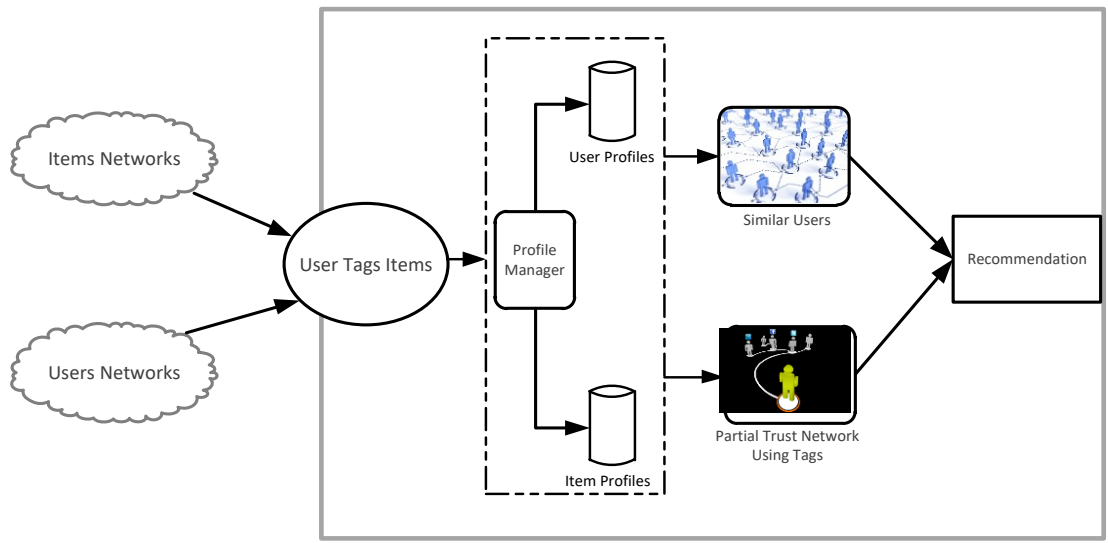

Fig. 1. Proposed model scheme – A collaborative filtering system works based on similarity and trust in a decentralized manner.

## 4.1. Profile Manager

Profile manager responsibility is to create or update user and item profiles when a user tags an item. It consists of two parts: User Profile Manager and Item Profile Manager.

### 4.1.1. User Profile Manager

User profile manager task is to create or update a user profile, when the user tags an item. In each user profile we store item URLs tagged by the user, tags assigned by the user to the item and the item weight that will be calculated for that user.

#### 4.1.1.1. Item weight

We want to fulfill the user-item matrix values [discussed in section 3.1] with item weights based on the tags each user assigns to different items, instead of using basic 0 and 1 values. In general, the more a tag has been used, the more interest the user has in the related resources. So we assume that users are more likely to prefer the resources bookmarked with the tags of high usage by them. And we also assume that a single user will use the same tags when describing similar resources to express his interests. Therefore we define tag weigh as follows.

$$W_{tag}(u,r) = \sum_{t_a \in tag(u,r)} W_{u,t_a} \quad (1)$$

Where tag(u,r) denotes the set of tags with which a user u has given to a resource r, $W_{u,t_a}$ denotes tag score of each tag $t_a$ in tag(u,r). Tag weight, $W_{tag}(u,r)$, measures how much a user u is interested in a tagged resource r, which implies a user's preference to for a resource.

Tag score, $W_{u,t_a}$, of each tag is calculated as follows.

$$W_{u,t_a} = \frac{freq(u,t_a)}{\sum_{l=1}^{k} freq(u,t_l)} \quad (2)$$

$freq(u,t_a)$ represents the times a tag ta has been used by a user u, k is the total number of tags a user u has tagged with any resources. According to the definition of tag score, tag weight takes a real number between 0 and 1, and the higher the weight is, the more interested a user is in a resource.

This approach does not suffer from natural language issues, such as ambiguity and synonym tags, which are the main problems when dealing with tags [29], since tag weight is generated within a single user's tag space. It goes without saying, each time a user applies a new tag in the system their profile will be updated, i.e. Wtag(u,r) values for different resources are dynamic and may change when a new resource is tagged.

### 4.1.2. Item Profile Manager

Item profile manager creates or updates an item's profile when a user tags an item. In each item's profile we store URLs of users who have tagged the item, list of tags each user has assigned to the item and the transaction trust value that we calculate for the related transaction. These values are dynamic and change with a new tag assignment. We define a transaction in a social tagging network as a triple (u,$T_{TR}$,r), denotes a transaction in the system for user u, annotating resource r with a set of tags $T_{TR}$.

#### 4.1.2.1. Transaction-level trust

**Trust:** If we take a deeper step towards the collaborative nature of social tagging systems, we will find that knowledge on the tagging scheme does exists in the way numerous users annotating resources with tags. Halpin et al. [30] have pointed out, if there are sufficient active users, over time, a stable distribution with a limited number of stable tags and a much larger "long-tail" of more idiosyncratic tags will develop. So we can consider this stabilized distribution as a convergence of the tagging scheme. We believe we can use the collaborative knowledge that implicitly exists in social tagging networks. In this work we estimate a degree of trust for users in two steps. In the first step a trust value is calculated for each transaction a user has within the system, based on comparing the assigned tags by the user, with those assigned by other users, which is the item profile manager task, which we will explain in this section. In the second step we shall find users trust values that will be discussed in section 4.4 in detail.

***Trust to a transaction:*** Information value or trust value of a transaction TR can be calculated as the average of the tags' tagging information values in $T_{TR}$.

$$V_{TR} = \frac{\sum_{t \in T_{TR}} V_r(t)}{|T_{TR}|} \qquad (3)$$

Where $V_r(t)$ is tag t's tagging information value.

$$V_r(t) = \frac{S_r(t)}{\sum_{t' \in T_r} S_r(t')} \qquad (4)$$

Where $S_r(t)$ is the number of times the tag t is used by separate users to annotate the resource r. A transaction with low information value indicates a divergence from crowd and a poor value of tagging information.

*4.2. Neighbors and Candidate Items*

Before we calculate similarity and user trust values in a decentralized approach we will define two terms, neighbors and candidate items that are needed for this purpose in the following.

*4.2.1. Neighbors*

Rating (or weighting) matrix in collaborative filtering systems has two demotions: user and item. This matrix is used to find similarities between users. If you take a look at the matrix you will realize that if the union of items between a user u with a user x is null, i.e. users u and x have not tagged any common item yet, similarity value between these two users is equal to 0. So in practice to make this matrix for user u in the item dimension we don't need users like x.

Therefor the neighbors of a user include users who have tagged at least one common item with the user. For this reason to specify the neighbors of a user u, we collect list of users that are recorded in profiles of items that u has already tagged (e.g. fig. 2).

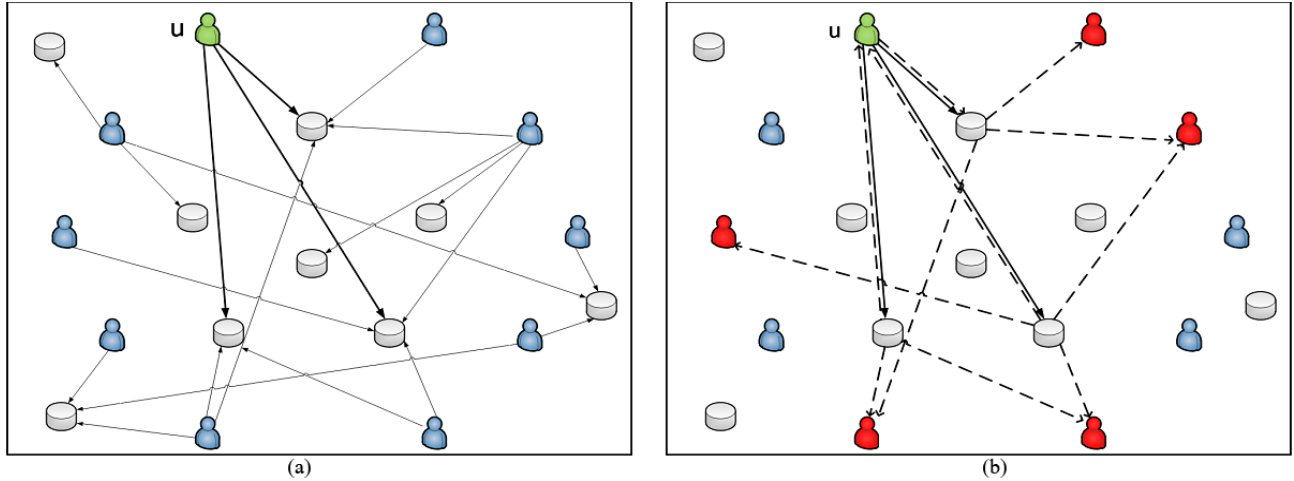

Fig. 2. An example of specifying neighbors of a user in a decentralized approach. (a) Shows a user u and other users and items tagged by each user in the system. (b) Shows how this system finds neighbors of the user u through items tagged by him.

### 4.2.2. Neighbors

As we mentioned, each user that has a nonzero similarity value with a user u, is counted as u's neighbor (e.g. fig. 3(a)). So the only users that participate in recommendation for user u are his neighbors, they represent the items already tagged by them, as candidate items that will be suggested to the user u. We define the union of all items that are recommended by a user's neighbors as candidate items (e.g. fig. 3(b)). It is obvious that we don't consider the set of items that are already visited by the user, as candidate items that will be recommended to them(marked by * in Fig. 3 (b)).

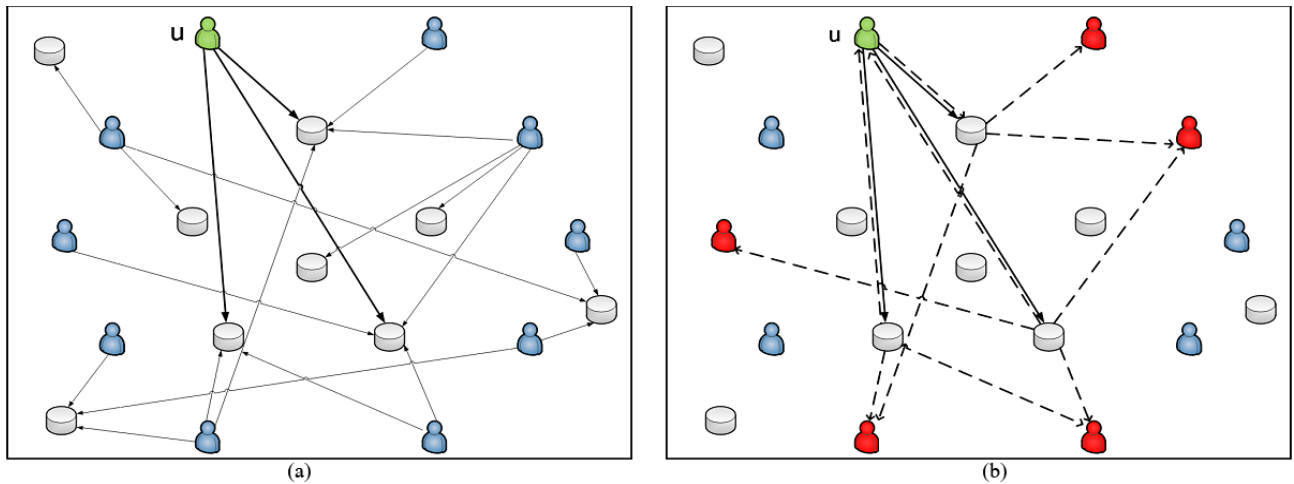

Fig 3. (a) Shows neighbors of a user which we specify in a decentralized approach. (b) Shows how we find candidate items using neighbors' profiles.

### 4.3. Similarity

User similarity calculation aims to mine the relationship among users. According to the approach we explained above, after we specify neighbors of a user and set of candidate items, we form user-item matrix with neighbors in user dimension and item dimension which is made up of candidate items and other resources that are tagged by the user. Matrix is filled with item weights which we keep in each item's profile. Since we don't need to collect all users and items information to make the weighting matrix to calculate similarities, this method leads to a noticeable size reduction in both dimensions compared to matrixes that are used in basic methods. There are a number of techniques to calculate the similarity between users, such as Pearson correlation coefficient, Spearman correlation coefficient, cosine similarity and Jaccard similarity. In this work we use cosine similarity method to compute the similarity value between user u and user v which is defined as follows.

$$sim\ (u,v) = \frac{\sum_{x \in X(u,v)} W_{tag}(u,x) \times W_{tag}(v,x)}{\sqrt{\sum_{x \in X(u,v)} W^2{}_{tag}(u,x)} \times \sqrt{\sum_{x \in X(u,v)} W^2{}_{tag}(v,x)}} \qquad (5)$$

Where X(u,v) is the set of resources that both user u and user v have tagged. $W_{tag}$ is the weight we calculated in equation 3.

### 4.4. User Level Trust

The straightforward idea is to use the average transaction information value, which we compute in equation 3 and store in related item profiles to represent a user trustworthiness, but this method does not take the resources importance into account. Since the popularity of resources in social tagging systems directly reflect the collaborative knowledge on resources quality, we define an importance value for each resource as follows,

$$I(r) = \frac{|U_r|}{|U_{r\prime}|} \qquad (6)$$

Where $U_r$ is the set of users who have tagged resource r and $U_{r\prime}$ is the set of users in the similarity matrix. Finally, we define a user u trust value as the weighted average of transaction information value, with resource importance as weight,

$$Trust_u(n) = \frac{\sum_{TR\prime \in TR_n} I(r) * V_{TR\prime}}{|TR_n|} \qquad (7)$$

### 4.5. Recommendation

In this section we use trust and similarity values to select best neighbors for item recommendation and then give our suggestions based on these users information.

***Selecting neighbors:*** As our objective is to provide a resource-recommendation model with both similarity and trust information, we combine the two weights into a single one in two steps. First, as the measured values of trust and similarity have different scales we normalize them to adjust them to a notionally common scale. For this purpose we find similarity values of the nearest and the farthest neighbors to a user u {Max$_u$(similarity), Min$_u$(similarity)} and by using min-max normalization (equation 8) we change the similarity values of neighbors so the nearest neighbor will get the value 1 and the farthest one value will become 0.

$$\forall n \in neighbor(u): Value_n = \frac{Sim(u,n) - Min_u(similarity)}{Max_u(similarity) - Min_u(similarity)} \quad (8)$$

In the same way we normalize trust values as follows.

$$n \in neighbor(u): T_n = \frac{Trus_u(n) - Min_u(Trust)}{Max_u(Trust) - Min_u(Trust)} \quad (9)$$

Where Max$_u$(Trust) and Min$_u$(Trust) are minimum and maximum trust values of user u's neighbors respectively.

In the second step we define the rank value for each neighbor n of user u as their total weight which is given in equation 10.

$$RankValue_u(n) = \lambda(Value_n) + (1 - \lambda)T_n \quad (10)$$

Where parameter λ is introduced to adjust the significance of the two normalized weights. In contrast to the traditional log-based ratings, our item-recommendation model uses both similarity and trust information to select better neighbors.

***Generating recommendations:*** Finally the recommendation module selects k neighbors with the highest rank values for resource suggestion. We use equation (11) to calculate each recommended item score,

$$score(u,r) = \frac{\sum_{v \in TopNeighbor(u)} W_{tag}(v,r) \times sim(u,v)}{|TopNeighbor(u)|} \quad (11)$$

Where |TopNeighbor(u)| denotes the number of top-n neighbors of a user u.

# 5. Experiments

*5.1. Data set*

The data set we have used in our experiments is the MovieLens data. In MovieLens site users can rate, tag and search for their favorite movies. This data set was published in May 2011 by GroupLens research group. It includes 2112 users, 10197 movies (include tagging and rating information), it has 13222 distinct tags and 47957 tag assignments (tas), i.e. tuples [user, tag, movie] to 5908 different movies. Average tas per user is 22.696 and distinct movies tagged by each user is 12.13 on average. The data set is divided into 2 parts: 20% of each user transactions form the testing set and the remaining 80% of transactions form the training set. We removed the movies that had not been bookmarked more than once, since they don't represent valuable information for us.

*5.2. Implementation and results*

We implemented our program with four different methods, The first set of experiments we conduct aims to analyze the impact of basic CF, it means similarity based in which user-item matrix is filled with 0 and 1 values. In the second set of experiments, we evaluate the impact of tag weight alone, i.e. similarity based model in which user-item matrix is filled with item weights to calculate similarities between users. In the third time we combined basic CF method with trust values estimated for neighbors of each user. Finally we evaluate our proposed model, the combination of similarity matrix (using item weighting approach) and the trust method we explained in section 4. We vary neighborhood sizes from 5 to 50 by an interval of 5 and recommended 10 top items to each user. To evaluate our results we used Recall, Precision and Coverage.

Recall and precision are used to evaluate system accuracy in top-10 recommended items for each user. Similar to Sarwar et al. [23], our evaluations consider any item in the recommendation set that matches any item in the testing set as a "hit". Note that we use mean average recall and mean average precision in our evaluations.

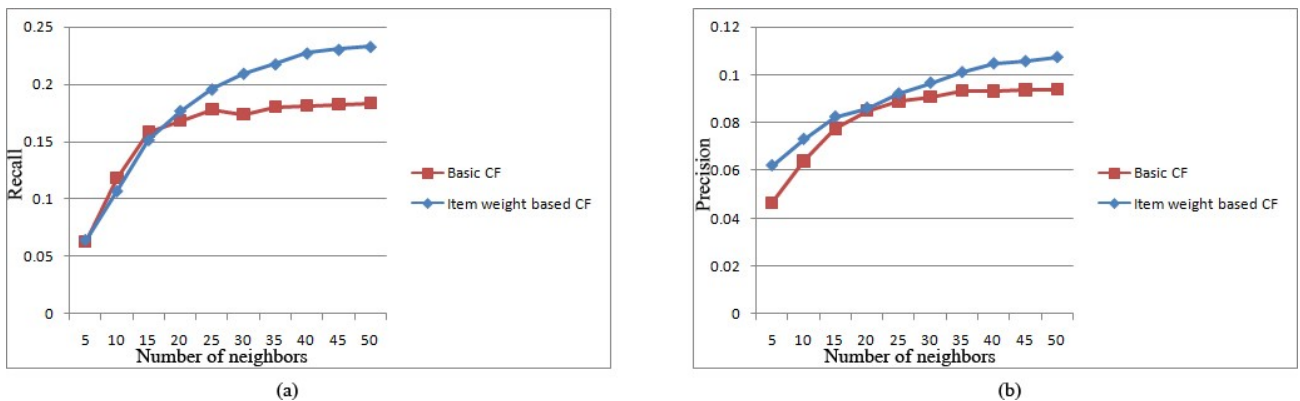

Fig. 4. (a) Recall and (b) precision results for the first and second experiments – Impact of item-weight strategy on recommendation quality.

Figure 4 shows the positives impact of item-weight strategy on precision and recall values. It is usually expected that recall decrease where precision increase and vice versa, but since we suggest only top 10 items to each user, both precision and recall accuracy get better when we increase the number of neighbors for recommendation process; so in the rest of this section we only show the results of calculated recalls for our different experiments.

The effect of trust-aware approaches (experiments 3 and 4) on recall are illustrated in Fig.5. These methods have led to better quality in recall for every number of neighbors compared to models based on only similarity values.

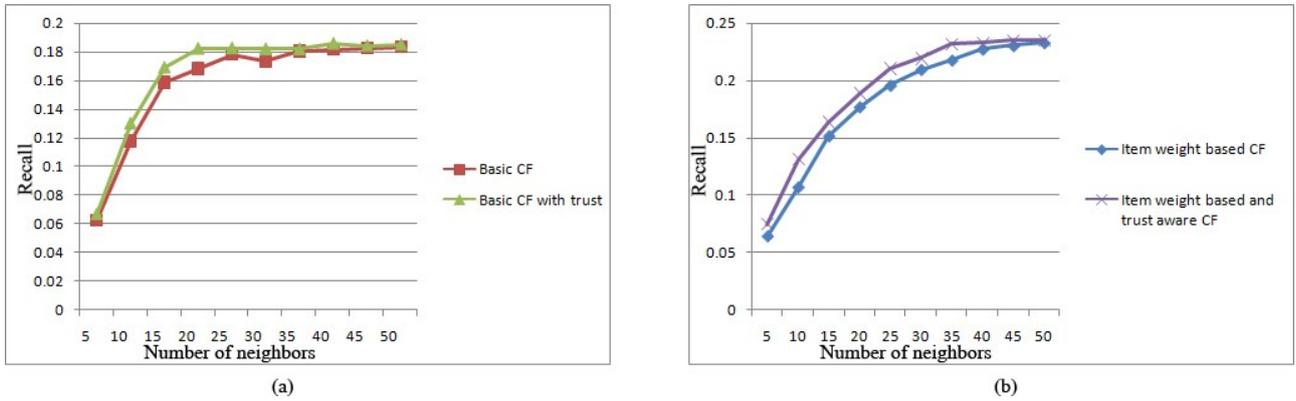

Fig 5. . Recall results - Impact of trust on recommendation quality.

Not surprisingly, our computational approaches with trust information outperform both traditional log-based (basic CF) model and item-weight model. As shown in Fig. 6.(a) weighting strategy has noticeable effect on recall results while trust strategy has led to slight improvement in accuracy of the recommender system.

Coverage is used for recommendation suitability [31], which measures the percentage of a dataset that the recommender system is able to provide predictions for; systems with lower coverage may be less valuable to users, since they will be limited in the decisions they are able to help with. Coverage is particularly important for the *Find All Good Items* task, since systems that cannot evaluate many of the items in the domain cannot find all of the good items in that domain. Fig. 6 (b) shows coverage results of the four implemented approaches.

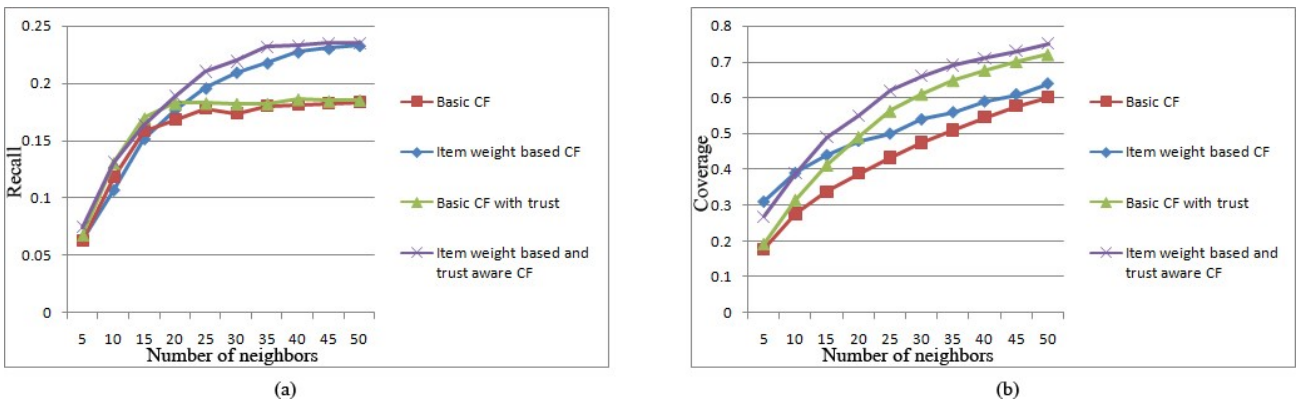

Fig. 6. (a) Recall results for 4 implemented approaches. (b) Coverage results for 4 implemented approaches

Comparing the results as we expected methods using weighted tags and items (experiment 2 and 4) have better accuracy than the methods which use basic user-item matrix for their similarity calculation. As it can be seen in coverage graph, there are significant improvements in trust based approaches. We can conclude from recall results that in the selected MovieLens data set there are not noticeable malicious users or spam tags, but in total, trust-aware models have led to selecting better neighbors, so we have improvement in finding all good items and the total recommended items set covers more elements in the testing set.

# 6. Conclusion and future works

We proposed a collaborative filtering system which uses trust beside similarity to select neighbors of users with more accuracy; it is a novel approach of calculating trusts based on users tagging behaviors alone. Trust in recommender systems is used to protect system against malicious users and noisy or spam tags which is one of the weaknesses of these systems. Another advantage of the proposed system is that it calculates user interests toward their bookmarked items independently and makes user-item matrix based on their interests. So the system does not suffer from some natural language problems such as synonym words or ambiguity in tags. The proposed model aslo leads to matrix size reduction which helps to lessen scalability issue in these kind of systems.

In our work, there is a limitation that we used only MovieLens dataset for experiments. It is needed to evaluate our computational approach using other datasets. Therefore, for the future work, we want to evaluate our item recommendation model with other data sets, such as Delicious, Flickr and Citeulike. We also plan to find other ways of aggregating trust with the standard collaborative filtering method and also to work on other weaknesses of these systems such as cold start users.